# A Simulation Method for Particle Fragmentation Based on Energy Landscape

Yupeng Jiang[1], Fernando Alonso-Marroquin[1], Hans J Herrmann[2], and Peter Mora[3]

1. School of Civil Engineering, The University of Sydney, Sydney NSW, Australia
2. PMMH, ESPCI, 17 quai St. Bernard, 75005 Paris, France
3. College of Petroleum Engineering and Geosciences, King Fahd University of Petroleum and Minerals, Dhahram 31261, Saudi Arabia

Abstract: We propose a method for the simulation of particle fragmentation based on the calculation of the energy landscape inside the particle. The landscape of strain energy is calculated in terms of internal stress using the principles of damage and fracture mechanics. Numerical calculation of the landscape's ridges is used to determine the breakage criterion as well as the shape of the post-breakage fragments. This method provides a physical-based understanding of the breakage effect in granular material.

Keywords: Particle breakage, Fragmenting model, Sub-particle stress, Granular material.

## 1. Introduction

The mechanical properties of granular materials are fully governed by its discrete components and the contact forces. Breakage of a single particle could be caused by the strong compressive loading exerted by its surrounding environment. Breakage is affected by particle shape [1, 2], the number of particles [3-5], and the distribution of contact forces [1, 6]. In granular materials, particle breakage is commonly observed, which has a major effect on the mechanical properties, such as density [6], and compressibility [7]. As a collective response to particle breakage, the particle size distribution (PSD) [3-7] and normal compression line (NCL) [4-7] are adopted as indicators of the breakage effect. Their relations with the state variables have been studied through experiments [1-5,7], which provides a solid foundation for understanding particle breakage. However, because of the difficulties of direct laboratory measurements, the relation between a single breakage event and the change of macro-scale mechanics is still poorly understood.

The particle replacement paradigm of the discrete element method (DEM), i.e. replacing the particle by its fragments when it reaches a breakage criterion, is used in many numerical models [8-14]. The existing methods possess high computational efficiency and the ability to reproduce PSD and NCL that agree well with experimental results [8]. Generally, two major attributes must be considered when developing a particle replacement model: a proper breakage criterion and a realistic approach for the determination of the geometry of post-breakage particles.

The existing breakage criteria are based on the averaged stress tensor. Particle breakage happens if the maximum principle stress or invariants of the stress tensor, or a linear combination of them, reaches a critical value. This criterion has been adopted by many DEM-based breakage simulations due to high computational efficiency. The averaged particle stress tensor $\sigma$ can be calculated as:

$$\boldsymbol{\sigma} = \frac{1}{V} \sum_{n=1}^{N} (\mathbf{x}_n^c - \mathbf{x}_n^p) \mathbf{F}_n . \tag{1}$$

The average stress is given in terms of the contact forces $\mathbf{F}_n$ that is exerted on the particle, the positions of the centre of mass $\mathbf{x}_n^c$, the contact location $\mathbf{x}_n^p$, the total number of contacts $N$, here call *coordination number*, and $V$ the volume of the particle. The stress tensor $\boldsymbol{\sigma}$ can be used for stress-based breakage criteria such as the von-Mises, the hydrostatic, or the maximum principal stress criterion. A typical use of Eq. 1 is to calculate the particle stress stemming from two diametrically opposite forces of magnitude $F$ in the $y$-direction. For a spherical particle with a radius $r$ the only non-zero stress component is $\sigma_y=(6F/r^2\pi)$. For a cylindrical particle with a radius $r$ and thickness $L$, the non-zero stress component is $\sigma_y=(2F/\pi r L)$. The commonly used particle breaking criterion assumes that the particle breaks when the maximum principal component of the stress reaches a critical value $\sigma_{cr}$. To evaluate this critical particle strength, the "smaller is stronger" principle is also applied. This is based on the experimental evidence [15-17] that the particle strength is higher when the particle is smaller, $\sigma_{cr} \sim d^{-b}$, where $b$ is a positive exponent [17, 18, 19].

Despite its computational efficiency, the averaged stress tensor oversimplifies the stress distribution inside the particle, which is also known as intra-granular stress [20] or sub-particle stress [21]. Overlooking the internal stress distribution causes a large deviation from the actual strength values obtained from finite element analysis and laboratory measurement [11, 12]. The averaged stress tensor also offers little information for determining the geometry of post-breakage particles. A better resolution of particle breakage can be obtained by evaluating the stress field inside the particle, here called "sub-particle" stress. The need for higher resolution breakage models is justified for the Brazilian test, where a cylindrical rock specimen is subjected by two opposite forces. The analysis of the sub-particle stress predicts a tensile failure at the centre of the particle given by $\sigma_t=(F/\pi r L)$, where $\sigma_t$ is the maximal tensile stress of the specimen [11, 21]. This result is similar to the stress calculated by the average stress method in the paragraph above, except for a factor of 2. However, the stress average technique predicts a compressive failure, while the Brazilian test shows tensile failure. These deficiencies motivate us to propose an advanced particle breakage criterion based on the sub-particle stress.

In this paper, we develop a particle breakage method that is fully governed by the sub-particle stress state. An advanced physical perspective for the particle breakage is first introduced based on the principles of Griffith fracture mechanics [22]. The relationship between damage initiation, the orientation of fragmenting lines, and the sub-particle stress is discussed. The breakage criterion, as well as the fragmenting method, are then defined using the strain energy landscape function proposed by Lemaitre and Chaboche [23, 24]. Results are rigorously analysed with the experimental data of existing studies. The main purpose of this study is to introduce an advanced perspective of particle breakage based on sub-particle stress.

The paper is organized as follows: the theoretical background and breakage model are presented in Section 2. The validation for the single-grain prediction of particle breakage test is provided in Section 3. The discussion and general conclusions are made in Section.4.

## 2. Breakage model

2.1 Damage-fracture model

In our model, Griffith's elastic fracture mechanics [22] is regarded as the basic model for crack propagation. The particle is assumed as a *continuum* material (Fig. 1) that contains evenly distributed *micro-scale* elliptical cracks. The orientation and length of these flaws are random. The size and density of cracks are low enough so that the continuum mechanics approach is still applicable at the macroscale. Thus, the local sub-particle stress and the strain can still be calculated based on linear elasticity.

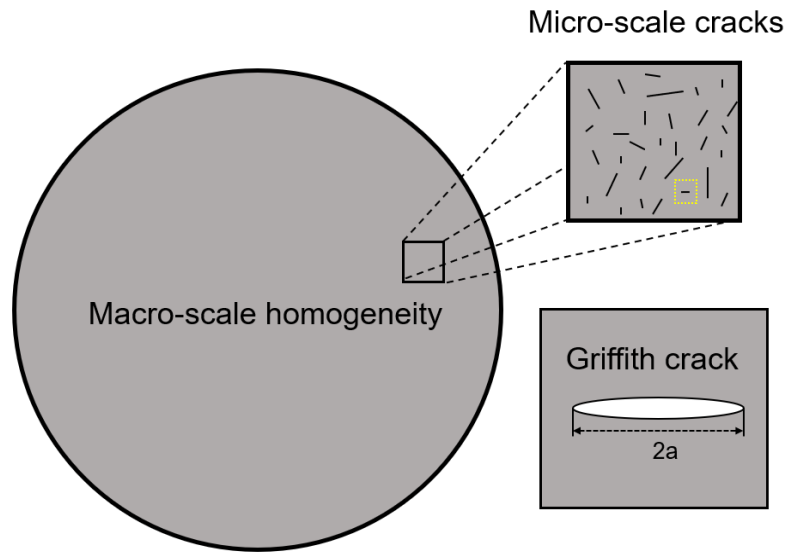

Figure 1. The material model for the particle at different scales.

A distribution of strain energy, here called energy landscape, is generated within the particle when external loads are applied. With the increase of strain energy, some local cracks are activated and start growing and coalescing with each other. Eventually, a group of activated cracks develops into a macro-scale fracture and causes splitting of the particle. This process describes the whole process of the single-particle breakage from the initial loading to the final breakage. The breakage criterion, the shape of replacement fragment, and the calculation of particle strength of new fragments will be defined based on this perspective.

In comparison to the existing theories, this physical perspective of breakage has several improvements. The breakage of the particle is regarded as a collective effect of crack growth [25] caused by the increase of strain energy. The size effect of particle strength, as an experimentally validated phenomenon, can still be incorporated into the theory since the maximum size of the crack is proportional to the particle size. The assumption of continuum material allows the efficient calculation of the sub-particle stress by linear elasticity. In this spirit, we do not consider the detailed properties of the micro-cracks and the dynamic progress of their propagation. In reality, the mechanical properties of the particle certainly change with the growth of the cracks. However, this progress occurs in a short time and over a small scale compared with the quasi-static deformation of the granular system.

2.2 Breakage criterion

Under a certain loading condition, the activation of a crack is fully determined by its size and orientation angle. The critical stress of the activation decreases with the increase of the half-side of the crack *a* as shown in Fig. 1 and varies with the orientation angle. For an equal crack size, the configuration that produces the largest stress at the tip of the crack defines the *critical angle* [26]. If the local strain energy is sufficiently high, the largest local crack in the critical angle will be the first to grow or be activated. This crack is defined as *critical crack* in this paper.

The energy landscape function $\psi(\sigma, v)$ depends on the magnitude of the local elastic energy, where $\sigma=\sigma(x, y)$ is the local sub-particle stress tensor and $v$ is the Poisson ratio. Similar to the von-Mises stress, the energy landscape function is a scalar value of the stress-based indicator for the elastic energy and independent of the material modulus. Now let us define a lower bound of $\psi$:

$$\varepsilon \leq \psi(\sigma, v),\qquad(2)$$

where $\varepsilon$ is a material parameter that serves as the lower limit for the activation of a local critical crack. It represents the threshold needed to generate damage in the particle; no local crack will start to grow if the value of $\psi$ is smaller than $\varepsilon$. Any area inside the particle that does not satisfy Eq. 2 can be regarded as an *intact domain*. If the local energy exceeds this limit the critical cracks are activated. The domain of the particle in this state is defined as the *damaged domain*. The contour line $\psi(\sigma, v)=\varepsilon$ is the boundary between the damaged and intact domains on the particle. The value of $\varepsilon$ governs the characteristic strength of the particle; if $\varepsilon=0$, the particle is damaged under any non-zero contact forces; if $\varepsilon=+\infty$, no damaged domain can be generated and the particle is unbreakable. Here, $\varepsilon$ is calibrated with the maximum tensile stress at the centre of the disc under diametrical loading.

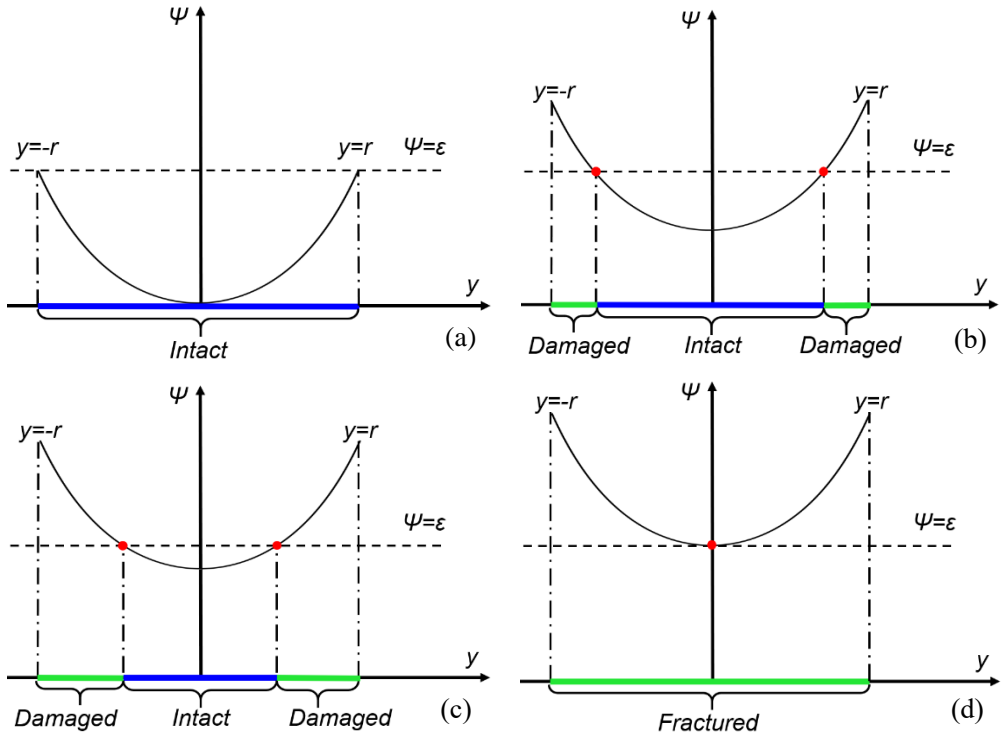

Figure 2. The variation of the energy landscape and its corresponding material state in one dimension. The value of $\psi$ is calculated along the central line ($x=0$) of the two-dimensional

Brazilian test with a radius of $r$. (a) no part of $\psi$ is above $\varepsilon$, the whole domain is intact. (b) the energy landscape increases and the domains with $\varepsilon \leq \psi$ are in the damage state; the red dots indicate the boundary between the damaged and intact state. (c) The value of $\psi$ further increases; the damaged domains expand and the intact domain shrinks. (d) the damaged domains cover the entire central line and breakage occurs.

The evolution of the energy landscape under increasing external load and the growth of the damage zones are illustrated in Fig. 2. Here we plot the energy landscape along the line connecting the two load points in the Brazilian test. With an increase of the external loads, the damaged domains will expand and eventually percolate through the intact domain. The red dots ($\psi=\varepsilon$) are the boundary between the intact segment and the damaged segments in Fig. 2 (b)-(d). We assume that the breakage occurs if the damaged domains cover the entire central line (Fig. 2 (d)). For the two-dimensional calculation, the red dots become a contour line that separates the damaged and intact domains. The particle breaks if the damaged domain contains a line that geometrically separates the particle. This line is regarded as the fragmenting line. In this way, the breakage criterion delivers the geometry of the new fragments through the sub-particle stress.

The optimal choice of $\psi$ is still an open question. Here we use the damage energy function of Lemaitre and Chaboche [23, 24]:

$$\psi(\boldsymbol{\sigma}, v) = \sigma_v \left[ \frac{2}{3}(1+v) + 3(1-2v)(\frac{\sigma_m}{\sigma_v})^2 \right]^{1/2}, \quad (3)$$

where $\sigma_v$ is the von-Mises stress and $\sigma_m$ is the hydrostatic stress. It is directly calculated from the Cauchy stress tensor. This equation describes the strain energy release that is associated with a unit damage growth. It considers both the influence of hydrostatic stress for brittle damage and of the von-Mises stress for plastic behaviour. Eq. 3 is used to determine the energy level for the local rupture or crack initiation in damage mechanics. Also, Eq. 3 is used as a comprehensive indicator of the density of deformation energy. It is similar to the energy release rate in linear fracture mechanics. The elastic energy release rate considers the effect of von-Mises stress and hydrostatic stress. Therefore, it is adopted as the elastic energy indicator for the activation of a local critical crack.

The calculation of sub-particle stress is relatively easy. It can be regarded as a boundary problem with a pure Neumann (traction) boundary condition. Eq. 4 states that the stress field inside the particle can be represented by a linear combination of fundamental solutions, the boundary traction **t** and the displacement **u** [21]:

$$\sigma_{ij}(\mathbf{x}) = \int_\Gamma D_{kij}(\mathbf{x}) t_k \mathrm{d}\Gamma - \int_\Gamma S_{kij}(\mathbf{x}) u_k \mathrm{d}\Gamma \quad , \ (i,j,k=1,2) \quad (4)$$

where the fundamental solutions $D_{kij}$ and $S_{kij}$ are given as:

$$D_{kij} = \frac{1}{r}\left\{(1-2v)[\delta_{ki}r_{,j} + \delta_{kj}r_{,i} - \delta_{ij}r_{,k}] + 2r_{,i}r_{,j}r_{,k}\right\}\frac{1}{4\pi(1-v)}, \quad (5)$$

$$S_{kij} = \frac{2}{r^2} \left\{ \begin{array}{l} 2\frac{\partial r}{\partial n}[(1-2\nu)\delta_{ij}r_{,k} + \nu(\delta_{ik}r_{,j} + \delta_{jk}r_{,i}) - 4r_{,i}r_{,j}r_{,k}] \\ +2\nu(n_i r_{,j} r_{,k} + n_j r_{,i} r_{,k}) + (1-2\nu)(2n_k r_{,i} r_{,j} + n_j \delta_{ik} + n_i \delta_{jk}) \\ -(1-4\nu)n_k \delta_{ij} \end{array} \right\} \frac{1}{4\pi(1-\nu)}, \quad (6)$$

where $r$ is the Euclidean distance between the boundary point and the inner point **x**; $\delta$ denotes the Kronecker delta function; **n** is the outward normal vector at the boundary point; and $r_{,i}$ denotes the calculation the first derivative of the corresponding variable $i$. The specific values of **t** can be obtained by dividing the contact forces by the size of their corresponding contact areas on the boundary, it serves as the initial boundary condition (Neumann condition). The displacement vector **u** of the whole boundary of the particle is unknown. It can be numerically obtained using the boundary element method [21]. After the displacement **u** is found, the Cauchy stress at any inner point **x** of the particle can be calculated by substituting **t** and **u** into Eq. 4.

Eqs. 5 and 6 are proportional to $1/r$ and $1/r^2$ respectively which indicates that the maximum deformation occurs at the point where the traction (source) is applied and decreases with the Euclidean distance from the source point. The strain energy particularly concentrates near the loading point and decreases quadratically with the Euclidian distance. It suggests that the damage domain should start from each source point and expand with an increase in the external loading. The fragmenting line can be defined once damaged domains that originate from the source points connect with each other and particle breakage occurs.

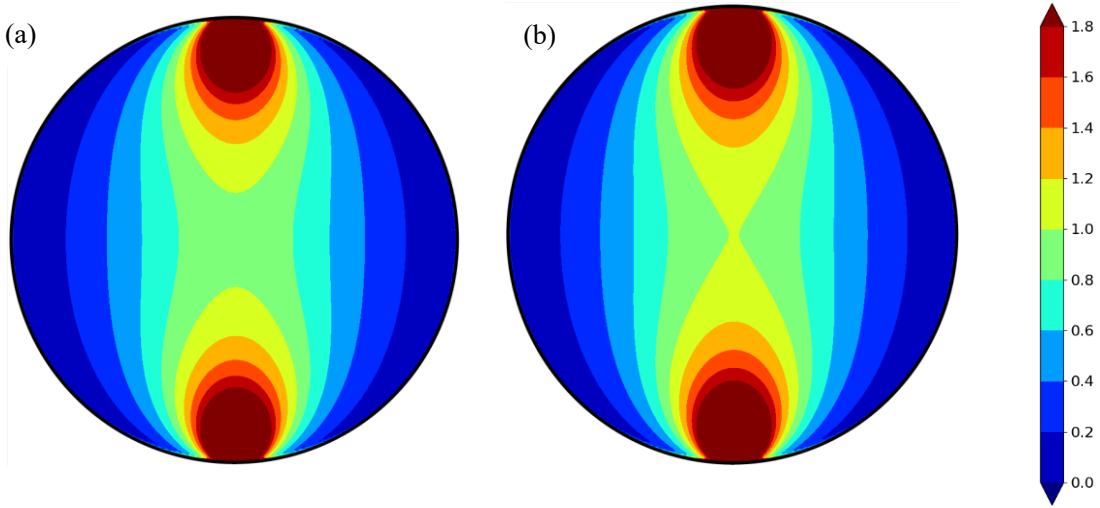

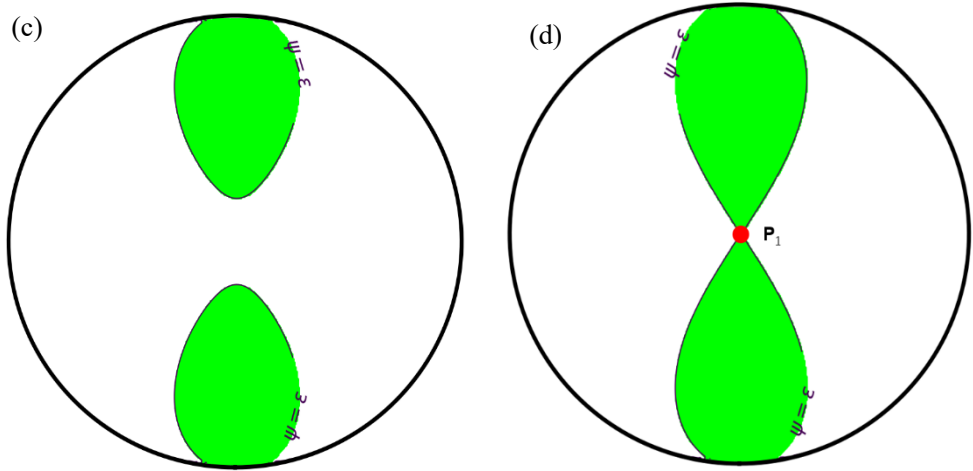

Figure 3. The growth and coalescence of the damaged domains with the diametrical loading in two dimensions. The red dot represents the connection point and the colour denotes $\psi$ as calculated from Eq. 3. The contour lines in (a) and (b) are the function $\psi$ normalized by $\varepsilon$. (a) shows the energy landscape of an insufficient loading; (b) the energy landscape of the breakage loading; (c) the boundary of $\varepsilon$ in (a) is not connected, (no breakage); (d) the boundary of $\varepsilon$ in (b) is connected due the increase of the contact forces so that breakage occurs.

The breakage criterion is shown in Fig. 3 based on Eq. 3; the damaged domain is illustrated in green. Similar to Fig. 2, the damaged domains are generated inside the particle and bounded by the intact area $\psi<\varepsilon$ in Fig. 3(c) since the force is not large enough. With the increase of the forces, the damaged domains expand and a line inside the particle can be defined when damaged domains that originated from opposite source points connect at the centre of the particle in Fig. 2(d). Such a path that has no intact area to block the growth and coalescence of the locally activated cracks inside the damaged domain; a macro-scale fracture will be formed. Based on these features, the breakage criterion for the diametrical loading is further defined by the existence of a *connection point* $(x, y)$ that satisfies:

$$\begin{cases} \varepsilon = \psi(\boldsymbol{\sigma}, v)\big|_{(x,y)} \\ \left\|\nabla \psi\big|_{(x,y)}\right\| = 0 \end{cases}, \qquad (7)$$

That is, the particle breakage occurs when the value of $\psi$ at the connection point $(x, y)$ equals $\varepsilon$, and the location of the connection point is a stationary point of the function $\psi$.

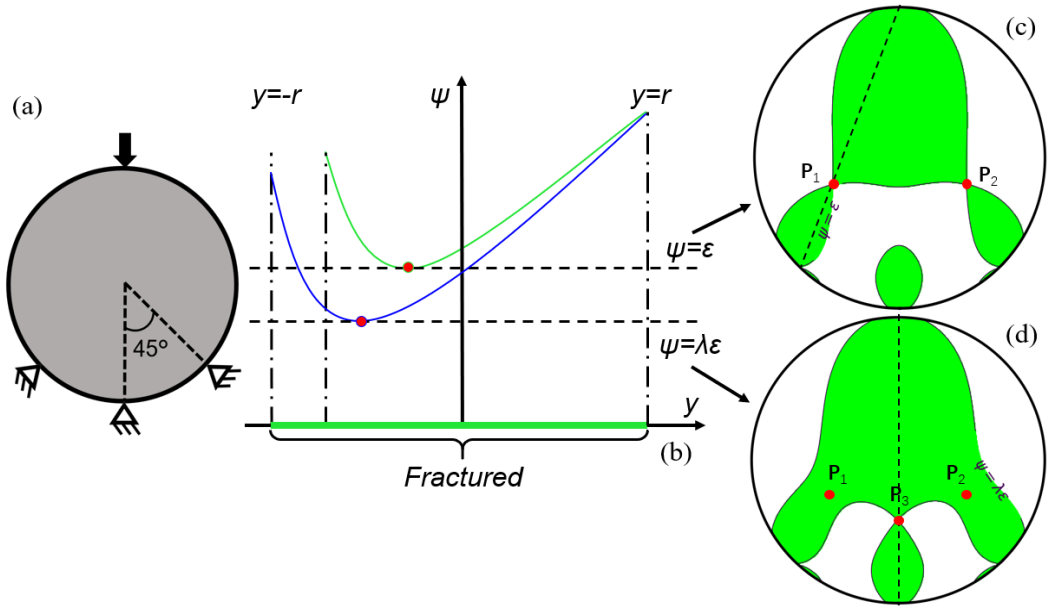

Figure 4. The growth and coalescence of the damaged domains with a high coordination number ($N$=4). (a) Load configuration. The black arrow represents the compressive force and the triangles denote the fixed boundary; (b) Projection of energy landscapes. The orange and blue lines represent the $y$-axis projected energy landscape along the dash lines shown in (c) and (d). (c) A damaged zone containing two connection points $P_1$ and $P_2$ resulting from $\psi=\varepsilon$; (d) Damaged zone containing all three connection points when $\psi=\lambda\varepsilon$.

The determination of the connection points becomes complicated if the number of load points, here defined as coordination number ($N$), is larger than two. It can be observed from Fig. 4 that the connection points increase as the $N$ increases. The values of $\psi$ at each connection points do not reach the threshold $\psi=\varepsilon$ simultaneously. This assumption is justified by the experiments of Salami et al [27]. Their experiment suggests that the particle breakage is a multi-stage process; the fragmenting lines are actually produced in a sequence before the equilibrium of the contact forces is broken by the generation of new fragments. In the particular case of the configuration shown in Fig. 4, the experiments show that the two breakage surfaces passing to the connection points $P_1$ and $P_2$ are created first. Then a secondary breakage surface passing to the point connection point $P_3$ is produced due to the damage of the primary fragments. Other experiments [2, 28] confirm this progressive damage pattern. Cracks and voids in the damaged area expand and therefore cause the deterioration of the material prior to the final breakage.

To account for this multi-stage fragmentation process, we assume that the particle reaches first the breakage criteria $\psi=\varepsilon$. Then the connection points $P_1$ and $P_2$ shown in Figure 4 (c) are calculated based on Eq. 7. These two points lead to two primary breakage surfaces. Then, it is assumed that the strength of the fragments is reduced by $\lambda\varepsilon$, where $0<\lambda<1$. Next, the damage domain is recalculated using $\psi=\lambda\varepsilon$, as shown in Figure 4 (d). If a small enough value of $\lambda$ is chosen, a third connection point $P_3$ satisfying $\psi(P_3)=\lambda\varepsilon$ is obtained. A lower value of $\lambda$ means a lower energy threshold, which may include more qualified connection points and produces more fractures. In Fig. 4 we use $\lambda=0.75$ to obtain the three connection points that ultimately define the three fracture surfaces observed in the experiments.

2.2 Fragmenting line

Here we propose a method to trace the fragmenting line based on the damage-breakage perspective and the energy landscape $\psi$. This method is governed by the idea that the fragmenting lines have the tendency to cross the area with the highest local energy. Based on this criterion, we calculate the ridges of the energy landscape using the Hessian matrix, which consists of the second derivatives of $\psi$ within the Cartesian coordinate system. The general direction of the fragmenting lines should correspond to the ridges of the energy landscape. This is underlined by the principle that the propagation and coalescence of micro-cracks have the tendency to stay within the area of high deformation energy. Therefore, the ridges of the energy landscape are adopted here as a first-order prediction of the fragmenting lines [29]. A ridge point is identified as a point where $\psi$ satisfies:

$$\frac{\partial \psi}{\partial u} = 0, \ \frac{\partial^2 \psi}{\partial u^2} < 0, \ \psi \geq \varepsilon, \tag{8}$$

where $u$ is defined as the direction of the eigenvector associated with the largest absolute eigenvalue of the Hessian matrix.

**3. Validation of single particle breakage**

The breakage scheme is validated with different values of coordination number on a two-dimensional Brazilian disc. The basic configuration of the six tests used in their paper [27] is shown in the first column of Fig. 5. The experiment used a cement disc as a sample in the Brazilian test. The active compressive load, which is illustrated through the black arrow, is applied at the top of the samples. The rest of the contact points are represented by linear clamping bars, which passively generate reaction forces under compression. Each loading case is named with a sequence of numbers. The first number denotes the coordination number and the next ones are the angles among the fixed contact points, The boundary element method is used to calculate the sub-particle stress. The simulation using 556 linear boundary elements. In the numerical simulation, the active and passive contacts are regarded as boundary compression and zero-displacement boundary condition respectively. The strength $\varepsilon$ is calibrated using the value obtained from the diametrical test.

Results of critical force and fragment shape are rigorously compared and analysed with the existing experimental data and images produced by Salami et al [27]. The results demonstrate that our sub-particle breakage analysis could provide valuable information about both breakage conditions and the shape of fragments.

3.1 Validation for the shape of fragments

The approximate shape of the fragment can also be retrieved from the numerical analysis of the energy landscape calculated from the sub-particle stress. The third row of Fig. 5 shows the damaged domains (orange colour) at the moment of breakage and the separation lines; the experimental images of the fractures are given in the last row of Fig. 5 for a direct comparison. It can be observed that the fractures obtained from the sub-particle breakage scheme agree well with the experimental images for each coordination number. Both the orientation and the number of fractures are properly captured. Especially, the pattern of fragmenting paths connecting different contact points is successfully predicted. It must be pointed out that the connection between a pair of contact points is a dominating feature for solid particle breakage, where the point-loading condition is common. This phenomenon has been confirmed by the experimental studies of Artoni et al [30] on a single-particle test. The result further validates our energy landscape model since it provides a proper explanation for the geometrical pattern of the fragmenting lines.

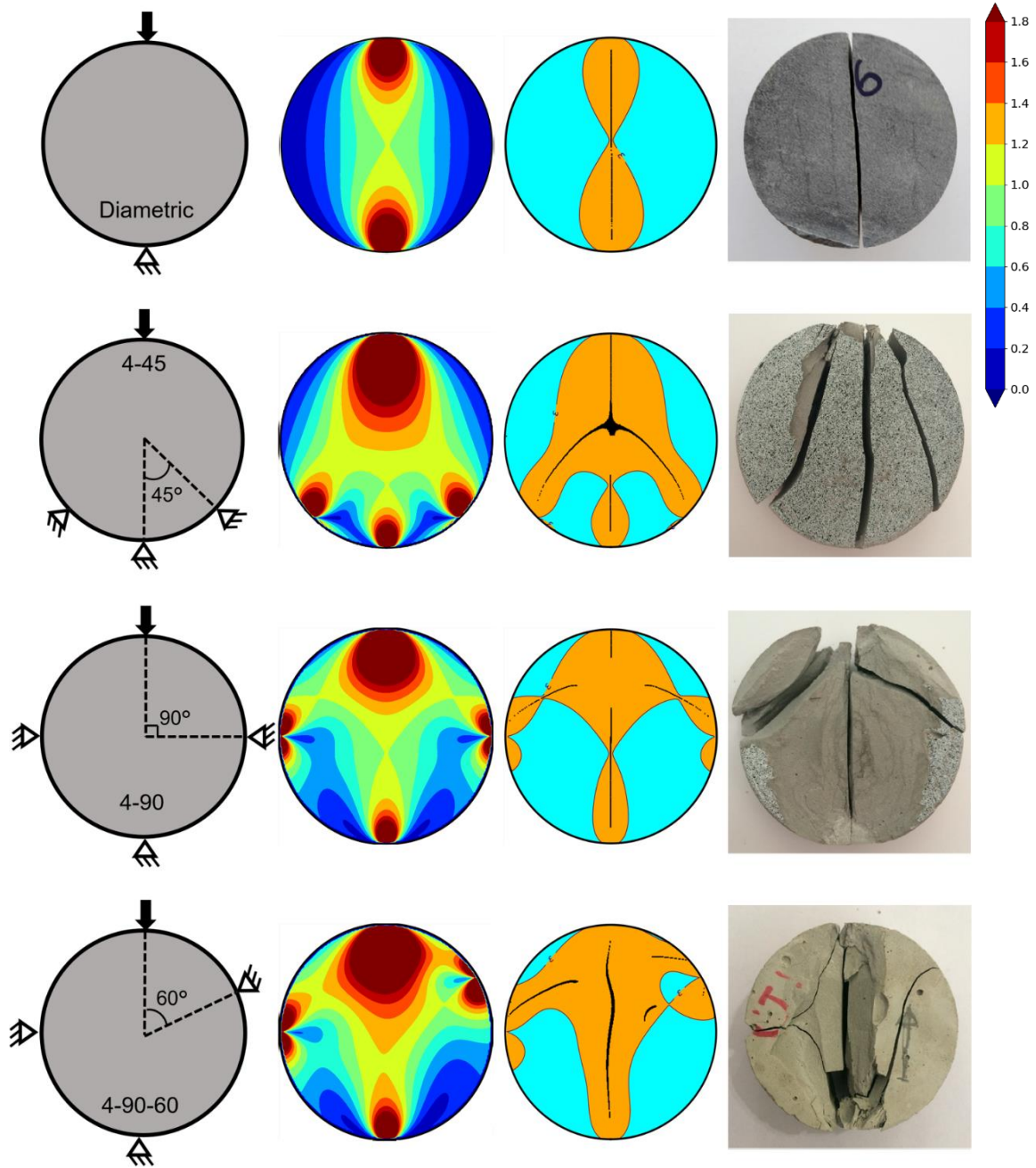

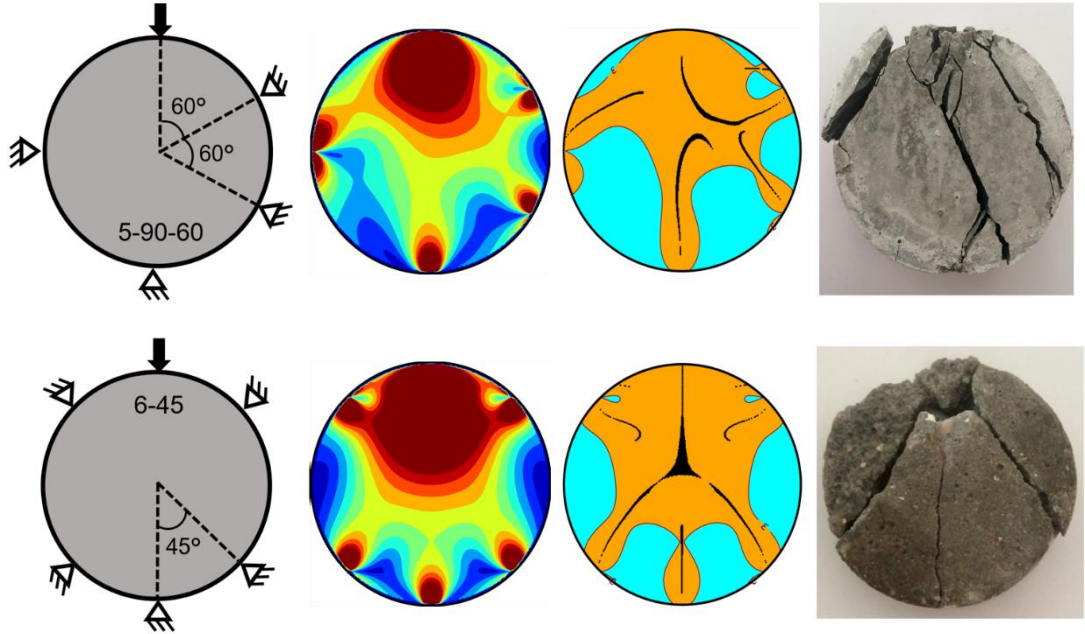

Figure 5. Comparison between the numerical and experimental results for the fractures of a circular particle; the first row shows the configuration of each test with different coordination numbers. The black arrow denotes the compressive force while the tailed triangle is the place where a zero-displacement boundary is applied; the second row contains the contour lines of $\psi=\varepsilon$ for different $\varepsilon$; the third row shows the numerical fragmenting lines provided by our method; the last row exhibits the images of the fragments from the experiment.

The agreement is particularly good for the cases of diametric, 4-90, 4-45 and 4-90-60. Minor differences are found in cases 4-90-60 and 5-90-60, where the numerical scheme provides fewer fragmentations than the experiment and have different shapes. Such differences are mainly due to the fact that our model does not consider the variations of the energy landscape that happen during the fracturing process. In reality, the fractures will interact and affect the generation and propagation of each other through the changes in the energy landscape. This process means that one fracture is able to attract, repel or arrest other fractures in its vicinity. Fracture propagation can in principle be captured by dynamically simulating the entire process, which is computationally expensive. In our method, the particle is treated as a whole entity before its final breakage and fragmentation. It approximates a strongly nonlinear process by a static energy landscape. The differences are acceptable considering the improvement in computational efficiency and the reduction of complexity brought by this method.

Other factors such as material heterogeneity [31] and the relative displacement between the sample and constraint points could also contribute to the disagreement between numerical and experimental results. Hence, minor differences are expected between the actual breakage and numerical simulation. The results indicate that such differences can be properly controlled to reach an optimized compromise of accuracy and computational efficiency. More importantly, the results further validate the assumption that the energy landscape is an important player of particle breakage.

3.2 Validation of the breakage criteria

The final breakage load for each case is defined as the critical force. The comparison of the critical forces at different coordination numbers is presented in Fig. 6. The analytical values in Salami et al.'s [27] study are calculated based on a formula proposed by Sukumaran et al [32], which includes an empirical nonlinear factor to approximate the effect of the coordination number on the critical force.

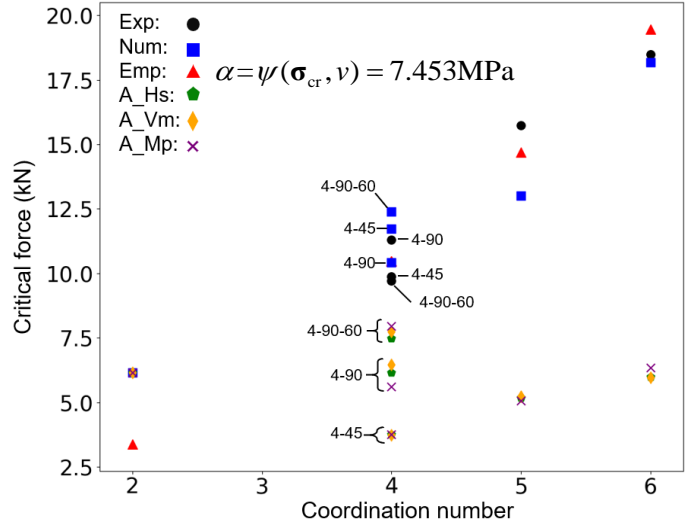

Figure 6 Critical force for different coordination numbers. The experimental results (Exp) are represented with black dots; the empirical results (Emp) are denoted with the red triangle and our numerical results (Num) are represented with blue squares. The results calculated from the averaged stress tensor are also provided for comparison, which includes hydrostatic stress (A_Hs), Von-Mises stress (A_Vm) and maximum principal stress (A_Mp).

It can be observed from Fig. 6 that each group of the critical force obtained from experimental, analytical and numerical tests increases with $N$. Such a result shows that the particle shows a stronger breakage resistance when it has more loading contacts. This is consistent with the conclusions made by several other studies about the effect of the coordination number. A rigorous analysis of this phenomenon is provided by Jiang et al [21] via a series of two-dimensional numerical studies of the sub-particle stress, which concluded that the increase of the coordination number can decrease the maximum principal stress and strain energy inside the particle. However, this study is only valid for a perfectly circular particle since the maximum principal stress increases with the coordination number for elliptical particles.

The results provided by our method show a rather good agreement with the experimental data. Our method is particularly useful for $N=2$ and 6, where the difference between the empirical formula and experiment data is distinctively large. The largest deviation of our method from the experimental data is 15% for $N=5$. The critical force calculated from the averaged stress fails to provides a proper approximation for all the cases of $N>2$. It cannot reflect the effect produced by the increase of the coordination number, which further proves the errors generated by using the averaged stress tensor are not negligible even for a circular particle.

Our results prove that our basic assumptions are valid for various loading situations. Both the empirical formula and our numerical method can be applied to approximate the breakage of circular particles under various coordination numbers. However, for the case of irregular particles, the validity of the empirical formula becomes questionable. The geometrical properties, such as void ratio and angularity, have a strong influence on the breakage resistance. It is difficult to generalize all the situations with one single empirical formula. Our numerical method, which is based on the sub-particle stress analysis and the deformation energy landscape, seems more suitable to provide the criteria for the breakage of irregularly shaped particles.

## 4. Conclusions

In this paper, we proposed a new physical perspective for particle fragmentation. This method overcomes two of the existing limitations of the use of the averaged stress tensor: the oversimplification of particle strength and the lack of sub-particle stress analysis. Our framework has unique advantages for understanding the fundamental mechanism of particle breakage at both individual particle and granular level. The energy landscape is derived from the theory of damage mechanics. The particle breakage is interpreted as the result of the percolation of the damage domains generated by the strain energy. The energy landscape of the particle under external loading is given by the sub-particle stress. The distribution of the damaged domain on the particle determines the breakage criterion and the geometry of the fragmenting lines.

The breakage limit and the geometry of the fragmenting surfaces are validated with the experimental results of the Brazilian test under various coordination numbers. The comparison indicates that both the critical force, calculated from the breakage criterion, and the shape of the fragments, agree well with the experimental data. In conclusion, our particle breakage theory unifies the breakage criterion, shape of the new fragments, and the sub-particle stress under one theoretical frame. It largely removes the use of empirical parameters and the problems caused by oversimplification. Therefore, the framework is expected to make a significant contribution to the study of the mechanism of particle breakage.

crushing strength using DEM. In: 5th World Congress on Particle. Orlando, FL: 1-8.